\documentclass[useAMS,usenatbib,onecolumn]{mn2e}
\usepackage{epsfig}

\def\ea{\ et al. \,}

\def\be{\begin{equation}}
\def\ee{\end{equation}}

\begin{document}

\title{Power Spectra of CMB Polarization by Scattering in Clusters}

\maketitle
\author{M. Shimon, Y. Rephaeli, S. Sadeh and B. Keating}

{M. Shimon$^1$, Y. Rephaeli$^{1,2}$, S. Sadeh$^2$ and B. Keating$^1$\\
\\
\\
1.\ Center for Astrophysics and Space Sciences, University
of California, San Diego, 9500 Gilman Drive, La Jolla, CA, 92093-0424\\ 
2.\ School of Physics and Astronomy, Tel Aviv University, Tel aviv, 69978, Israel\\} 
                                                                
\date{Feb. 12, 2009}

\begin{abstract}

Mapping CMB polarization is an essential ingredient of current cosmological 
research. 
Particularly challenging is the measurement of an extremely weak B-mode polarization 
that can potentially yield unique insight on inflation. Achieving this objective 
requires very precise measurements of the secondary polarization components on 
both large and small angular scales. Scattering of the CMB 
in galaxy clusters induces several polarization effects whose measurements can 
probe cluster properties. Perhaps more important are levels of the statistical 
polarization signals from 
the population of clusters. Power spectra of five of these polarization components 
are calculated and compared with the primary polarization spectra. These spectra 
peak at multipoles $\ell \geq 3000$, and attain levels that are unlikely to 
appreciably contaminate the primordial polarization signals.
\end{abstract}


\section{Introduction}

Primordial seed perturbations in the densities of radiation and matter, 
and the excitation of gravitational waves during an epoch of cosmological 
inflation, left their imprints on the cosmic microwave background (CMB). 
Generic models of inflation predict gaussian density fluctuations, 
and therefore all the statistical information characterizing the 
primordial CMB is expected to be encoded in the 
two-point 
correlation function defined on the celestial sphere, or equivalently in 
its harmonic transform - the angular power spectrum. Various processes 
that took place at later times when the evolution on small scales was 
nonlinear, can induce non-gaussian anisotropy. An important example is 
the anisotropy induced by Compton scattering of the CMB by hot gas in 
galaxy clusters - the SZ effect. Indeed, for this effect higher order 
spectra, e.g. the tri-spectra, may contain additional valuable information. 
Nonetheless, a wealth of information can be gleaned from the angular power 
spectrum itself. This work focuses on {\it power spectra} of CMB polarization 
induced by scattering in clusters.

While the primary temperature anisotropy, velocity-gradient-induced 
E-mode polarization, and gravitational-wave-induced B-mode polarization 
dominate the lowest and intermediate multipoles of the power spectra, 
the latter is damped on scales smaller than the horizon 
at recombination, 
and the former are exponentially suppressed beyond $l\approx 1000$ due to 
photon diffusion damping (Silk 1968). The latter process took place during 
recombination, when the radiation decoupled from the baryons for the first 
time. On these few arcminute scales (determined by the thickness of the last 
scattering shell), secondary signals induced by clusters of galaxies peak at 
multipoles, $l$, characteristic of galaxy clusters ($l\approx 2000-3000$). 
This secondary anisotropy gauges the formation and evolution of clusters; 
as such it depends on different combinations of the cosmological parameters 
than the primary CMB anisotropy. Its steep dependence on the basic quantities 
that determine the evolution of the large scale structure (LSS) - such as 
the critical density for collapse and mass variance - makes it particularly 
valuable for removing some of the inherent degeneracies in the cosmological 
parameters.

The generation of temperature anisotropy by clusters can be either through 
lensing by their gravitational field, which is clearly dominated by the dark 
sector of the cluster total energy (i.e. dark matter, and possibly also - 
indirectly - dark energy), or by Compton scattering of the radiation by 
intracluster (IC) 
gas, the main compoent of the baryonic sector. Cluster-induced lensing of 
the temperature and polarization anisotropy was the subject of several recent 
studies. It merely redistributes the anisotropy on the sky while conserving 
the total power. Lensing couples fluctuations of temperature or polarization 
on cluster scales to the lensing deflection angle, which can be directly related 
to the transverse gradient of the cluster projected gravitational potential. 
Therefore, lensing smoothes features in the power spectrum on scales comparable 
to the characteristic deflection angle. Lensing of the CMB by clusters also 
converts the polarization modes from E to B, which is indeed part of the full 
lensing signature of the LSS. 

In this paper we focus on the statistical polarization signals induced 
collectively by scattering in clusters. Temperature anisotropy is induced 
when the CMB is scattered by moving electrons; fully random motions give rise 
to the {\it thermal} SZ effect, and additionally also to the {\it kinematic} 
SZ effect when the cluster has a finite radial velocity in the CMB frame. The 
kinematic effect is typically an order of magnitude smaller than the thermal 
component. Additional information can in principle be extracted from the 
polarization state of the radiation, which probes other combinations of 
cosmological and cluster parameters. 

Various polarization effects were studied in the original work of 
Sunyaev \& Zeldovich (1980), and further elaborated upon by Sazonov \& 
Sunyaev (1999) and e.g, Audit \& Simmons (1999). Conversion of the 
primordial E-mode polarization to B-mode by the LSS has been 
considered by e.g. Seljak \& Zaldarriaga (2000), Lewis \& King (2006) 
and by Hu, DeDeo \& Vale (2007). Gravitational lensing of the primary 
CMB mixes the two E \& B polarization which makes their separation from 
the primordial polarization extremely challenging, necessitating the use 
of higher order statistics to de-lens the sky. While these delensing 
methods work well for the LSS (because they were optimized for this case), 
similar methods should be applied to remove cluster signals; however, this 
need may be circumvented if the scattering-induced polarization signals are 
very weak, as seems to be the case based on results presented in this paper.

Cluster-produced polarization signals are typically much smaller than 
the temperatore anisotropy by orders of magnitude. The only two such 
polarization components discussed so far in the context of their {\it 
statistical} imprint on the polarization power spectrum are those resulting 
from the primordial quadrupole and the quadrupole anisotropy associated with 
the transversal (Doppler) component of the bulk motion of the cluster (Cooray, 
Baumann \& Sigurdson 2005). The latter is second order in the cluster
transverse velocity, and linear in the optical depth. Similarly, 
we expect the quadrupole induced by tensor perturbations to generate 
polarization, but this effect is much smaller, as implied from the 
current upper limit, $0.22$ (95\% CL), on the tensor-to-scalar ratio 
(Hinshaw et al. 2008)

Two other relevant polarization components whose statistical properties 
were not studied arise from {\it double} scattering in the same cluster 
(Sunyaev \& Zeldovich 1980, Sazonov \& Sunyaev 1999). First scattering 
induces temperature anisotropy either by the thermal or kinematic SZ 
effect. If this temperature anisotropy contains a local quadrupole 
moment, polarization is induced upon second scattering. 
For a typical Thomson optical depth of IC gas, 
$\tau \sim 0.01$, these $\tau^2$-depenedent components are clearly very 
weak, with the thermal effect sourcing the largest of the two. We ignore 
here the minute signals due to radiation scattered during aspherical 
collapse of protoclusters, scattering in a moving cluster in which the 
radiation develops anisotropy due to lensing (Gibilisco 1997), or 
polarization produced by IC magnetic fields (Ohno et al. 2003) and 
relativistic magnetized plasma (Cooray, Melchiorri \& Silk 2002).

The main purpose of this paper is to address the polarization signals 
discussed above in the context of their {\it statistical} signature, i.e. 
the power spectrum, especially in light of the anticipated detection of 
lensing-induced B-mode signal by Planck and other experiments. Our 
calculated power spectra include the primary Poissonian contribution; 
for simplicity we ignore the smaller contribution due to angular 
correlation between clusters (Komatsu \& Kitayama 1999). The latter 
peaks on larger angular scales that reflect the correlation distance 
between clusters.

In section 2 we briefly review the basics of cluster-induced polarization. 
Calculations of the power spectra are outlined in Section 3, followed by 
results in Section 4, and a brief discussion in Section 5.

\section{Polarization Induced by Scattering in Clusters}

Scattering of the CMB by IC gas changes the radiation temperature along 
lines of sights to the cluster. In the non-relativisitc limit (to lowest 
order in gas temperature) the {\it thermal} SZ effect (Sunyaev \& Zeldovich 
1972) constitutes a fractional temperature change 
\begin{eqnarray}
\frac{\Delta T}{T}&=&yg(x)\\
g(x)&=&x\coth(x/2)-4\nonumber\\
y&\equiv &\int\sigma_{T} n_{e}\frac{kT_{e}}{m_{e}c^{2}}dl
\end{eqnarray}
where $x=h\nu/(kT)$ is the dimensionless frequency, $n_{e}$ and 
$T_{e}$ are the electron number density and temperature, $y$ is the 
comptonization parameter, $\sigma_T$ is the Thomson cross section, 
and the integration is along the line of sight. Hereafter we use $\Theta$ 
for the dimensionless gas temperature $kT_{e}/(m_{e}c^{2})$.

The second, closely related and smaller kinematic SZ effect is proportional 
to the line of sight (los) velocity of the cluster, $v_{r}$, and is 
independent of frequency, 
\begin{eqnarray}
\frac{\Delta T}{T}&=&-\int\sigma_{T}n_{e}\beta_{r}dl\nonumber\\
\beta_{r}&\equiv &\frac{v_{r}}{c}.
\end{eqnarray}

Compton scattering can polarize incident radiation if it has a 
quadrupole moment. The CMB has a global quadrupole moment and a 
non-vanishing quadrupole moment is induced by scattering in the 
cluster. The degree of linear polarization and its orientation are 
determined by the two Stokes parameters
\begin{eqnarray}
Q&=&\frac{3\sigma_{T}}{16\pi}\int n_{e} dl \int \sin^{2}\theta\cos
2\phi T(\theta,\phi) d\Omega\nonumber\\
U&=&\frac{3\sigma_{T}}{16\pi}\int n_{e} dl\int \sin^{2}\theta\sin
2\phi T(\theta,\phi) d\Omega,
\end{eqnarray} 
where $\theta$ and $\phi$ define the relative directions of 
the incoming and outgoing photons, $d\Omega$ is an element of 
integration over the solid angle and $T(\theta,\phi)$ is the 
temperature of the incident radiation; we use temperature-equivalent units. 
Since the los is taken to be along the z-axis for convenience, 
the angles $\theta$ and $\phi$ are actually defined with respect to the 
outgoing photon in this system. The average electric field defines the 
polarization plane with a direction given by
\begin{eqnarray}
\alpha=\frac{1}{2}\tan^{-1}\frac{U}{Q}
\end{eqnarray}
and the total polarization (which is the quantity of interest to us here) 
is defined as
\begin{eqnarray}
P\equiv\sqrt{Q^{2}+U^{2}}.
\end{eqnarray}

Another relevant process is the E-B mixing by gravitational 
lensing. This effect can convert parity-even (E-mode) polarization 
to parity-odd (B-mode) polarization (Zaldarriaga 2001). 
In fact, the largest B-mode signal 
induced by galaxy clusters is due to the gravitational lensing of the 
primary E-mode polarization. Since CMB lensing by the LSS was extensively 
studied in the past decade, and the power spectrum of CMB polarization 
due to lensing is readily obtained with Boltzmann codes, we do not elaborate 
on this conversion here (Zaldarriaga \& Seljak 1998).
   
In the following subsection we describe the polarization generated by Compton 
scattering when a quadrupole moment is induced by electrons moving either at 
the cluster peculiar veocity or thermally. When the global quadrupole moment 
of the CMB is taken explicitly into account, then polarization induced by 
scattering in the gas is treated in the limit when the gas is viewed as `cold', 
i.e. the second order correction due to random electron motion is ignored.

\subsection{Polarization of Fully Isotropic Incident Radiation}

Scattering of the CMB in a cluster at rest in the CMB frame results in 
local anisotropy due to the different pathlengths of photons arriving 
from various directions to a given point. This anisotropy provides the 
requisite quadrupole moment; second scatterings then polarize the 
radiation. If the cluster is not resolved, no net polarization would be 
measured. Nonetheless, it is useful to explore the signal associated 
with double scattering since it is expected to dominate over the other 
polarization signals in {\it rich} clusters (Shimon, Rephaeli, O'Shea \& 
Norman 2006) for which $\tau^{2}$ is not negligibly small. 

\begin{figure*}
\epsfig{file=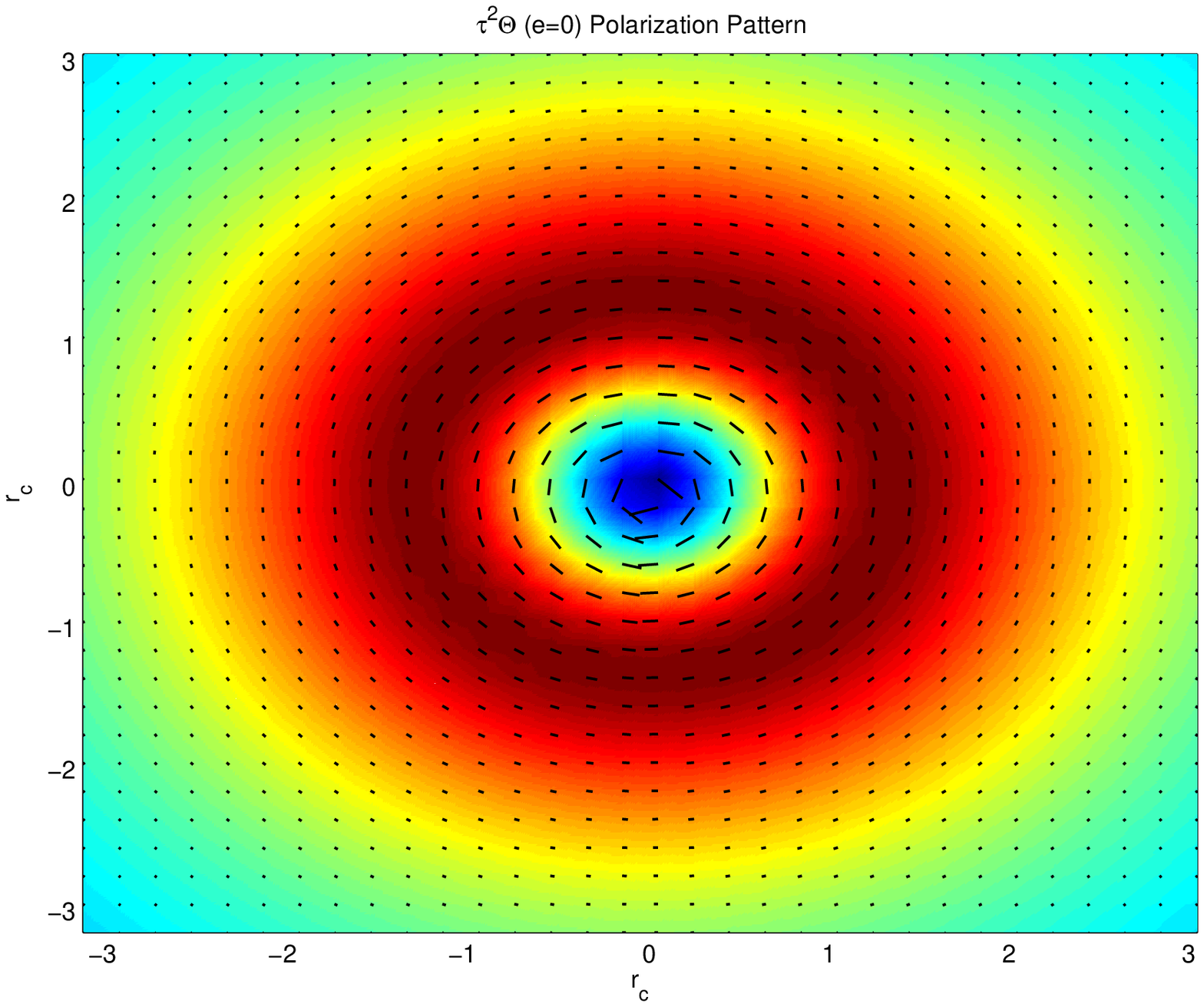, height=7cm, width=8cm,clip=}
\epsfig{file=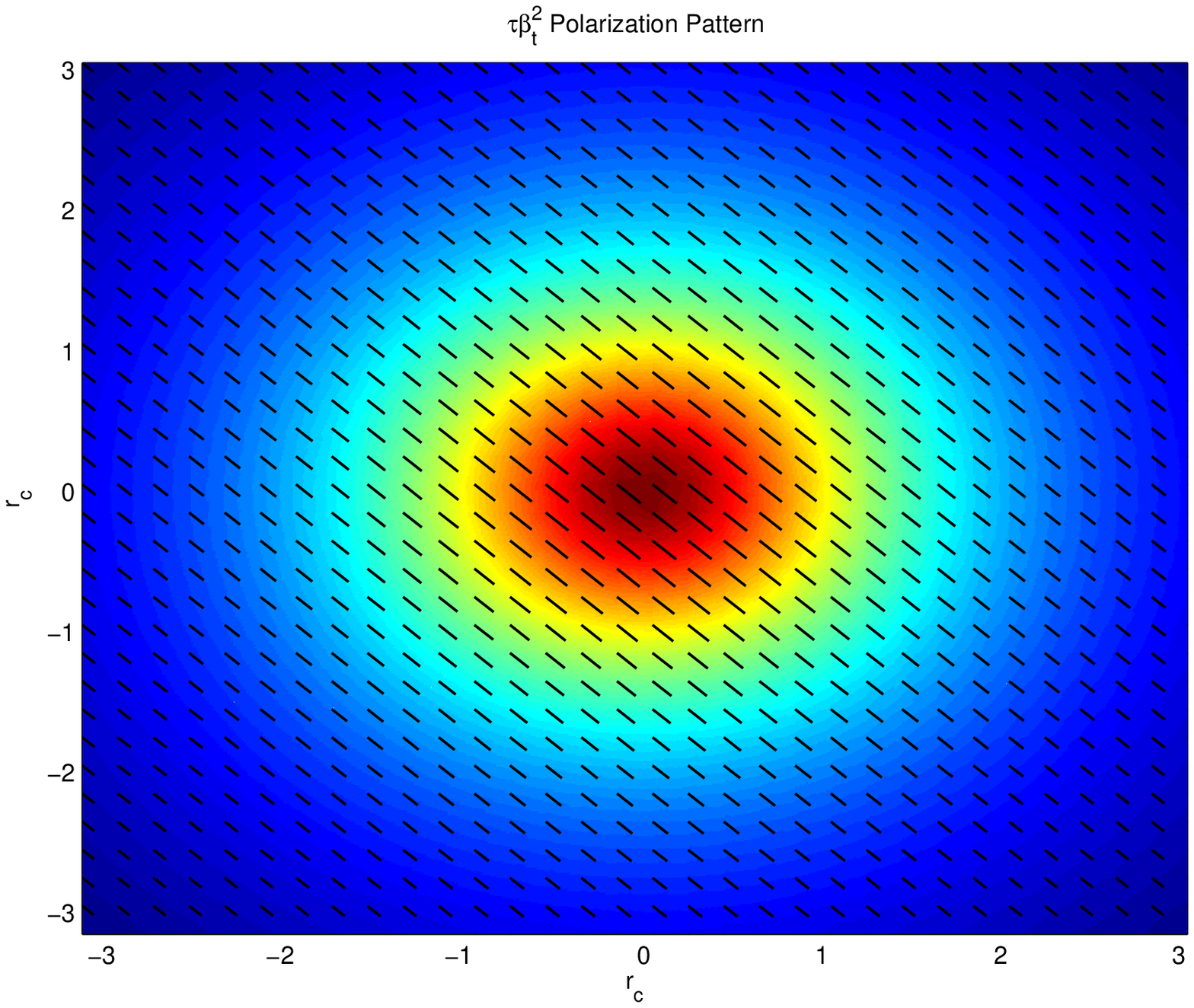, height=7cm, width=8cm,clip=}
\caption{SZ polarization patterns of the double scattering $\tau^{2}\Theta$ 
($e=0$, $x>x_{0}$) and $\tau\beta_{t}^{2}$ 
(cluster moving south-west to north-east) effects.}
\end{figure*}

The CMB appears anisotropic in the frame of a non-radially moving cluster; 
scattering by IC electrons then polarizes it. Two polarization components 
are induced; the first is linear in the cluster velocity component 
transverse to the los, $v_t \equiv\beta_{t}c$, but quadratic in 
$\tau$; the second is linear in $\tau$ but quadratic in $\beta_{t}$. 
The spatial patterns of the various polarization components can be 
readily determined when the gas distribution is spehrically symmetric. 
The polarization patterns arising from scattering off thermal electrons 
are isotropic in a spherical cluster (Figure 1, left panel) while the 
corresponding patterns of the kinematic components are clearly anisotropic 
due to the asymmetry introduced by the direction of the cluster motion 
(Figure 1, right panel). More realistic gas distributions, substructure 
and high internal velocities result in complicated polarization patterns 
(Lavaux, Diego, Mathis \& Silk 2004, Shimon, Rephaeli, O'Shea \& Norman 2006).

\subsubsection{The $\tau^{2}\beta_{t}$ and $\tau \beta_{t}^{2}$ 
Components}

The degree of polarization induced by double scattering in a 
non-radially moving cluster was determined by Sunyaev \& Zeldovich 
(1980) in the simple case of uniform gas density,
\begin{eqnarray}
P=\frac{1}{40}\tau^{2} \beta_{t}.
\end{eqnarray}
This polarization component is frequency-independent in equivalent temperature 
units since it is a first order Doppler shift. A more complete 
calculation of this and the other polarization components was formulated 
by Sazonov \& Sunyaev (1999). Viewed along a direction 
$\hat{\bf{n}}=(\theta,\phi)$, the temperature anisotropy at a point 
$(X,Y,Z)$, $\Delta T(X,Y,Z,\theta,\phi)$, leads to polarization upon second 
scattering. The Stokes parameters are calculated from Equation (4), 
\begin{eqnarray}
Q(X,Y)\pm iU(X,Y)&=&\frac{3\sigma_{T}}{16\pi}\int dZ n_{e}(X,Y,Z)\int d\Omega\sin^{2}(\theta)e^{\pm 2\phi}\Delta T(X,Y,Z,\theta,\phi)\nonumber\\
&=&\frac{3\sigma_{T}}{4\pi}\int dZ n_{e}(X,Y,Z)\int d\Omega Y_{2}^{\pm 2}(\theta,\phi)\Delta T(X,Y,Z,\theta,\phi)
\end{eqnarray}
$Y_{2}^{\pm 2}$ are the $l=2$ and $m=\pm 2$ spherical harmonics. The 
temperature change resulting from first scatterings is 
\begin{eqnarray}
& &\frac{\Delta T(X,Y,Z,\theta,\phi)}{T}=\sigma_{T}
\int d\vec{l}(X',Y',Z',\theta,\phi)n_{e}(X',Y',Z',\theta,\phi)\hat{n}\cdot
\beta(X',Y',Z'),
\end{eqnarray}
and the optical depth through the point $(X,Y,Z)$ in the direction 
$(\theta,\phi)$ is 
\begin{eqnarray}
\tau(X,Y,Z,\theta,\phi)=\sigma_{T}\int n_{e}(X',Y',Z')d\vec{l}
(X',Y',Z',\theta,\phi).
\end{eqnarray}
$Q(X,Y)$ and $U(X,Y)$ fully describe the linear 2D polarization field.

To study the polarization profile we define the following polar coordinates
\begin{eqnarray}
X&=&r\cos\psi\nonumber\\
Y&=&r\sin\psi 
\end{eqnarray}
where $r$ is the radial distance from the cluster center. A photon 
travels a distance ${\bf l}$ before scattering
\begin{eqnarray}
X'-X&=&l\sin\theta\cos\phi\nonumber\\
Y'-Y&=&l\sin\theta\sin\phi\nonumber\\
Z'-Z&=&l\cos\theta.
\end{eqnarray}
In these coordinates
\begin{eqnarray}
\frac{\Delta T(X,Y,Z,\theta,\phi)}{T}
=\sigma_{T}\int dl(X',Y',Z',\theta,\phi)n_{e}(X',Y',Z',
\theta,\phi)\beta(X',Y',Z').
\end{eqnarray}
Assuming the gas velocity is constant (i.e., internal bulk velocities are 
relatively low) and that the gas density has the familiar $\beta$ profile 
with index $\beta=2/3$, such that the electron density is  $n(r)=n_{0}/(1+r^{2})$, 
where the radial distance from the center is now taken to be in units of the core 
radius, $r_c$, we obtain 
\begin{eqnarray}
\frac{\Delta T(X,Y,Z,\theta,\phi)}{T}
=\sigma_{T}n_{0}r_{c}\int_{l=0}^{\infty}\frac{dl}{1+[r^{2}+
z^{2}+2l(r\sin\theta\cos(\psi-\phi)+z\cos\theta)+l^{2}]}
\end{eqnarray}
where both $l$, and $z$ are given in units of $r_{c}$. Writing 
\begin{eqnarray}
& & A^{2}\equiv 1+r^{2}+z_{0}^{2}\nonumber\\
& & B\equiv r\sin\theta\cos(\psi-\phi)+z\cos\theta 
\end{eqnarray}
and integrating over $l$ yields 
\begin{eqnarray}
\frac{\Delta T(X,Y,Z,\theta,\phi)}{T}
=\sigma_{T}n_{e}^{0}r_{c}\beta\left[\frac{\pi}{2}-
\tan^{-1}\left(\frac{B}{\sqrt{A^{2}-B^{2}}}\right)
\right]/\sqrt{A^{2}-B^{2}}. 
\end{eqnarray}
Inserting this into Eq.(8) and carrying out the 3D numerical integration 
over $z$, $\theta$ and $\phi$ we obtain $Q$ and $U$ as functions of 
$r$ and $\psi$ only. The {\it total} polarization is then obtained from 
Eq.(6) $P=\sqrt{Q^{2}+U^{2}}$, and by virtue of the fact that $P$ is a 
scalar quantity, independent of $\psi$. By fitting $P(r)$ over a large 
range of values $r$ (e.g. $0<\frac{r}{r_{c}}<10$) we obtain the following 
simple expression
\begin{eqnarray}
P(r)=\frac{3}{16\pi}(n_{e}^{0}\sigma_{T}r_{c})^{2}\beta
\left(\frac{ar}{1+br+cr^{2}}\right) 
\end{eqnarray}
where the constants $a$, $b$ and $c$ are obtained by fitting to 
the numerical 3D integrations described above
\begin{eqnarray}
a&=&0.57938006\nonumber\\
b&=&-0.85163944\nonumber\\
c&=&0.58122608.
\end{eqnarray}

The second kinematic polarization component is $\propto \tau\beta_{t}^{2}$; 
this component is generated by virtue of the fact that the radiation appears 
anisotropic in the electron frame if the electron motion has a nonvanishing 
transversal component. From the definition of the Stokes parameters Eq.(4) 
and the fact that the Doppler effect depends on the angle between the photon 
direction and the velocity vector, and by choosing the $Z$ axis to coincide 
with the direction of the electron velocity, one obtains (Chandrasekhar 1950).
\begin{eqnarray}
\frac{d\sigma}{d\Omega}=\frac{3\sigma_{T}}{8}
\left(1-\mu_{0}^{2}\right)P_{2}(\mu'_{0})\, ,
\end{eqnarray}
where $P_{2}(\mu'_{0})$ is the second Legendre polynomial, the 
expression for the $Q$ parameter of the scattered radiation is 
\begin{eqnarray}
Q(\mu)=\frac{3}{8}\tau(1-\mu_{0}^{2})
\int_{-1}^{1}P_{2}(\mu'_{0})I(\mu'_{0})d\mu'_{0}\, .
\end{eqnarray}
Here, $\mu'_{0}=\cos\theta'_{0}$, $\mu_{0}=\cos\theta_{0}$, 
are the cosines of the angles between the electron velocity and the incoming and 
outgoing photons, respectively. Expanding the apparent angular 
distribution of the radiation in Legendre polynomials, and keeping terms 
up to $\beta^2$, the quadrupole moment is determined. The polarization 
of the singly scattered radiation is then calculated (Sazonov \& Sunyaev 
1999) by using Equation (4). The level of this polarization component 
(Sunyaev \& Zeldovich 1980) is 
\begin{eqnarray}
Q=\frac{x^{2}e^{x}(e^{x}+1)}{20(e^{x}-1)^{2}}
\tau\beta_{t}^{2} .
\end{eqnarray}
In our chosen frame of reference the Stokes parameter $U$ vanishes due 
to azimuthal symmetry. Therefore, the total polarization amplitude, $P$, 
is equal to $Q$ and the polarization is orthogonal to $\beta_{t}$. 
Relativistic corrections (to the non-relativistic expression of Sunyaev 
\& Zeldovich 1980) were calculated by Challinor, Ford \& Lasenby (2000) 
and Itoh, Nozawa \& Kohyama (2000). These corrections generally amount to a $\sim 10\%$ 
reduction in the value of Q, and are therefore neglected in our calculations.

\subsubsection{The $\tau^{2}\Theta$ Component}

Analogous to the $\tau^{2}\beta$ component discussed above, double 
scattering off electrons moving with random thermal velocities can induce 
polarization that is proportional to $\tau^{2}\Theta$.
The anisotropy introduced by single scattering is the thermal component 
of the SZ effect with temperature change $\Delta T_{t}$. Its dependence 
on frequency is contained in the following analytic approximation to the 
exact relativistic calculation (Itoh, Kohyama \& Nozawa 1998, Shimon \& 
Rephaeli 2004)
\begin{eqnarray}
\frac{\Delta T(X,Y,Z,\theta,\phi)}{T}
=\sigma_{T}
\sum_{i=1}^{5}F_{i}(x)\int n_e({\bf r})\Theta({\bf r})^{i}dl. 
\end{eqnarray}
In Equation (22) the integration is along the photon trajectory prior to 
the second scattering, $F_{i}(x)$ are spectral functions of $x$ 
(Shimon \& Rephaeli 2004). The polarization is obtained by inserting 
this expression in Eq.(8) and repeating the procedure described in 
section (2.1.1). 

The polarization patterns of these two effects are illustrated in 
Sunyaev \& Zeldovich (1980) and Sazonov \& Sunyaev (1999); as can be 
seen from Figure 1, the polarization due to double scattering on the 
thermal plasma is expected to vanish when averaged over a beam due 
to its circular or radial symmetry. When high resoultion, sub-arcminute 
experiments will be operational the effect will be measured, if sensitivity 
reaches the few nK level and foreground removal and systematics can be 
controlled at the required level.

While typical clusters are not perfectly spherical and some residual
signal will survive the beam convolution, this signal is expected to 
be very weak; only second order in the cluster ellipticity $O(e^{2})$. 
Typically, the correction due to ellipcity is comparable to or smaller 
than few percent.

\subsection{Polarization of Anisotropic Incident Radiation}

The large scale anisotropy of the CMB includes a cosmological
quadrupole moment at the level of $\approx 15\mu K$, as measured by the all-sky 
surveys of COBE and WMAP. Knowing that the probability for generating 
polarization by scattering in clusters is a fraction $\sim\frac{\tau}{10}$ of 
the incident quadrupole, and that $\tau \sim 0.01$, we expect the resulting 
polarization signal to be $\approx$15 nK, therefore detection would be extremely 
challenging, even with next generation experiments. 

A similar but smaller component results from scattering of the radiation 
with a quadrupole moment generated by gravitational waves (from 
tensor perturbations). Its level is likely to be much smaller; the current 
limit on the tensor-to-scalar ratio is $T/S<0.4$, deduced from analysis 
of CMB measurements. The upper (95\%) confidence limit from WMAP alone is 
$0.43$ (Dunkley et al. 2008) and $0.22$, if the analysis 
includes also SN and BAO (Komatsu et al. 2008). We describe 
the polarization induced by the primordial quadrupole (both scalar 
and tensor contributions) below. The main difference between the two is that 
while density waves do grow during cosmic evolution, tensor perturbations 
do not, and since we sum the contributions from the entire population of 
clusters, the redshift evolution of the population has to be known reasonably 
accurately.

\subsubsection{Scalar Contribution}

The global quadrupole moment is imprinted on the CMB by the Sachs-Wolfe 
(SW) effect at the surface of last scattering, and is further boosted by 
the evolving gravitational fields of density inhomogeneities - the 
integrated Sachs-Wolfe (ISW) effect. There is no ISW contribution in 
a purely matter dominated universe, but around recombination 
the expansion was still partially driven by the residual radiation 
(which constituted about 1\% of the energy budget), and more recently 
(at $z<0.3$) the expansion became dominated by dark energy, which 
again caused decay of the gravitational potential. Scattering by IC 
electrons polarizes the radiation at a level proportional to the product 
of the rms of the primary quadrupole moment and the cluster scattering 
optical depth (Sazonov \& Sunyaev 1999). Using the WMAP normalization 
of the CMB quadrupole moment (Bennett \ea 2003), the maximal polarized 
signal is expected to be $\simeq 2.6\tau \, \mu K$, and its 
all-sky average is $\sim 60\%$ of this value (Sazonov \& Sunyaev 1999). 

It has been noted that the dependence of this polarization component on 
the CMB quadrupole moment could possibly be used to reduce cosmic variance 
(Kamionkowski \& Loeb 1997), but this seems doubtful (Portsmouth 2004). 
Another suggestion is that the dependence can probe dark energy models 
through the redshift evolution of the quadrupole. We note that the 
polarization induced by the CMB quadrupole can be distinguished from 
other cluster-induced polarization components by virtue of its large-scale 
distribution, reflecting that of the primary CMB quadrupole (e.g Baumann 
\& Cooray 2003), and the fact that it is independent of frequency (when 
expressed in temperature units). The usually quoted value of the quadrupole 
moment $Q_{rms}\approx 15\mu K$, refers to the primordial quadrupole 
(generated at recombination) but this quadrupole is directly affected by 
the evolving gravitational potential, and since perturbations in this potential 
can possibly alter the quadrupole, it has a redshift dependence imprinted by 
the evolving potentials. For a power law primordial power spectrum of density 
perturbations $P(k)\propto k^{n}$, the rms value of the quadrupole is 
(e.g. Hu 2000)
\begin{eqnarray}
Q^{2}_{{\bf 
rms}}(z)&=&\frac{5}{48}\delta_{H}^{2}(1+z)^{2}D^{2}(z)\Omega_{m}^{2}
(d_{A}H_{0})^{1-n}\Gamma_{{\bf sw}}(n)\nonumber\\
\Gamma_{{\bf sw}}(n)&\equiv &3\sqrt{\pi}\frac{\Gamma[(3-n)/2]\Gamma[(3+n)/2]}
{\Gamma[(4-n)/2]\Gamma[(9-n)/2]}
\end{eqnarray}
where $\delta_{H}=4.2\times 10^{-5}$, the growth function is
\begin{eqnarray}
D(z)=\frac{H(z)}{H_{0}}\int_{z}^{\infty}dz'(1+z')
\left(\frac{H_{0}}{H(z')}\right)^{3}/\int_{0}^{\infty}dz'(1+z')
\left(\frac{H_{0}}{H(z')}\right)^{3},
\end{eqnarray}
and the 
Hubble function is
\begin{eqnarray}
H(z)/H_{0}=\sqrt{\Omega_{m}(1+z)^{3}+\Omega_{\Lambda}},
\end{eqnarray}
(where $H_{0}$ is the present Hubble constant). 
In the case of a flat power spectrum $\Gamma_{{\bf sw}}(n=1)=1$ where $n$ 
is the spectral index of scalar metric perturbations, $\Gamma(x)$ is the 
Gamma function, $\Omega_{m}$ and $\Omega_{\Lambda}$ are the 
matter and vacuum density today in units of critical density, and 
$d_{A}$ is the angular diameter distance to redshift z. The 
polarization level in temperature units is (Sazonov \& Sunyaev 1999) 
\begin{eqnarray} 
P=\frac{\sqrt{6}}{10}\tau Q_{{\bf rms}}^{{\rm scalar}}.
\end{eqnarray}

\subsubsection{Tensor Contribution}

The scalar metric perturbations dominate the temperature anisotropy at 
least on small scales but there is also an observational upper limit on 
the tensorial contribution on large scales ($l<100$), as well as 
theoretically expected upper limits (i.e. the latter is $r<0.3$ for 
the simplest inflationary 
models). It is sufficient to gauge this small tensor contribution by 
the tensor-to-scalar ratio $r\equiv T/S$. The overall normalization 
should come from observations, and indeed it turns out that the 
fractional scalar perturbations are of order $\sim 10^{-5}$. Current 
upper limits from CMB temperature anisotropy experiments should be 
considered weak upper limits, for if inflation indeed occurred following 
a phase transition in the grand unification (GUT) era, a much lower value, 
$r\approx 0.01$, would be expected. It should be noted that, due to lensing 
of the CMB by the LSS,which peaks at $l\approx 1000$ but still leaks to the 
angular degree scales (where the primordial tensor perturbations peak)
there is a lower limit to the level of tensor-to-scalar ratio 
($r\approx 0.001$) which can be inferred even from ideal measurements 
(with noise-free detectors) of the primordial B-mode. However, inflation 
could have taken place before or after the GUT era. The ultimate test will 
be measurements of B-mode polarization, or a direct detection of the stochastic 
gravitational waves generated during inflation (presently considered extremely 
unlikely).

As we show below, similar to the polarization induced by the scalar 
SW and ISW effects, we expect a small polarization signal to be induced 
by scattering, and since the effect scales as 
$C_{l}^{P}\propto\tau^{2}C_{2}^{{\rm tensor}}$, and the 
current upper limit is $r<0.3$, we expect this signal to be at most 
$\sim 30\%$ of its scalar counterpart. This component contains both E 
and B modes (due to the nature of gravitational waves)  
and has the potential to assist in tightening the limits on the 
amplitude of gravitational waves. Also, as mentioned above, 
unlike the scalar quadrupole, 
the quadrupole induced by gravitational waves does not evolve 
significantly with redshift, i.e. the ISW effect for the tensor modes 
effectively vanishes. It is sensitive only to the anisotropic component 
of the energy-momentum tensor, and in fact decays by about 10\% due to 
neutrino streaming on quadrupole scales (Weinberg  2004), which is a small effect, 
ignored here. This yields $P=\frac{\sqrt{6}}{10}Q_{{\rm 
rms}}^{{\rm tensor}}$ as in the case of polarization induced by 
Compton scattering of the scalar quadrupole moment. Again, 
$Q_{{\rm rms}}^{{\rm tensor}}$ is obtained from 
$Q_{{\rm rms}}^{{\rm scalar}}$ (Eq. 23) by setting 
$n\rightarrow n-1$ and the growth function, which depends on redshift, is 
set to 1, followed by multiplying the resulting power spectrum by $r$, 
the current upper limit on the tensor-to-scalar ratio.

\section{Power spectrum}

Although the temperature anisotropy and polarization induced by galaxy 
clusters are intrinsically non-gaussian and their angular power spectra 
do not contain all the statistical information, these spectra still 
provide important information on the level of the anisotropy and its 
characteristic scales. The {\it angular} power spectrum is essentially 
the projection of the {\it processed} 3D power spectrum on the 2D 
celestial sphere. There are two characteristic angular scales in the 
problem; the angular size of a cluster and the typical angular 
correlation angle on the sky between neighboring (sufficiently rich) 
clusters. The first is typically a few arcminutes and the correlation 
angle is $\sim 1^{\circ}$ (if only rich clusters are considered) 
which correspond to $l\approx 2000-3000$ and $l\approx 200$, respectively. 
Their relative importance depends on the corresponding 3D power spectra 
which consists of the Poisson and correlation (clustering) terms. 
The contribution due to correlations is typically an order of magnitude 
smaller (e.g., Cooray, Baumann \& Sigurdson 2005) and is not considered 
here. The power spectrum of the anisotropy due to the population of clusters is
\begin{eqnarray}
C_{l}=\int\int |\tilde{\xi}(l,M;z)|^{2}\frac{dn(M;z)}{dM}\frac{dV}{dz}dMdz
\end{eqnarray}
where $\tilde{\xi}$ is the Fourier transform of the polarization 
generated in each cluster
\begin{eqnarray}
\tilde{\xi({\bf l})}=\int d^{2}\theta\xi(\theta)e^{i{\bf l}\cdot{\bf\theta}}.
\end{eqnarray}

The above integral is over the mass function of clusters, for which the 
Press-Schechter distribution (or one of its variants) is usually adopted
\begin{eqnarray}
\frac{dn(M;z)}{dM}=-F(\mu)\frac{\rho_{b}}{M\sigma}\frac{d\sigma}{dM},
\end{eqnarray}
where
\begin{eqnarray}
F(\mu)=\sqrt{\frac{2}{\pi}}e^{-\frac{\mu^{2}}{2}}\frac{\mu}{\sigma_{R}},
\end{eqnarray}
$M(R)\equiv\frac{4\pi}{3}\rho_{b}R^{3}$ and the mass variance is
\begin{eqnarray}
\sigma^{2}(M;z)=D^{2}(z)\int\frac{dk}{k}\frac{k^{3}P(k)}{2\pi^{2}}|W(kR)|^{2}.
\end{eqnarray}
Here $\mu\equiv\delta_c(z)/\sigma(z)$ is the critical overdensity 
for collapse in terms of the mass variance at redshift $z$, and
$\rho_b$ is the background density at $z=0$, and 
$W(q)=\frac{3}{q^{3}}[\sin(q)-q\cos(q)]$ is a top-hat window 
function. The processed density fluctuation power spectrum 
$P(k)\equiv Ak^{n}T^{2}(k)$ is given in terms of the transfer 
function $T(k)$ (which is specified below). The normalization constant 
is determined in terms of the observationally deduced value of $\sigma_8$, 
the mass variance on a scale of $8h^{-1}$ Mpc. 

The mass variance evolution with redshift is given by 
\begin{eqnarray}
\sigma(M;z)=\frac{g[\Omega_{m}(z)]}{g[\Omega_{m}(0)]}\frac{\sigma(M;0)}{1+z}  
\end{eqnarray}
where
\begin{eqnarray}
\Omega_{m}(z)=\Omega_{m}(0)(1+z)^{3}. 
\end{eqnarray}
For the function $g[\Omega_{m}(z)]$ we use the approximate expression 
(Carroll, Press \& Turner 1992)
\begin{eqnarray}
g[\Omega_{m}(z)]=\frac{2.5\Omega_{m}(z)}{\Omega_{m}(z)^{4/7}-\Omega_{\Lambda}
+(1+\Omega_{m}(z)/2)(1+\Omega_{\Lambda}/70)}.
\end{eqnarray}
We adopt the standard CDM transfer function
\begin{eqnarray}
T(k)=\frac{\ln(1+2.34q)}{2.34q}[1+3.89q+(16.1q)^{2}+(5.46q)^{3}+(6.71q)^{4}]^{-1/4}
\end{eqnarray}
with $q\equiv k/(\Omega_{m}h^{2}) Mpc^{-1}$ (Bardeen et al. 1986).

The virial relation is used for the gas temperature 
is obtained assuming virialization
\begin{eqnarray}
kT_{e}=\frac{GM\mu_{H}m_{p}}{3\beta r_{c}(M;z)}
\end{eqnarray}
where the core-radius is obtained from the spherical collapse model
\begin{eqnarray}
r_{c}(M;z)=\frac{r_{0}}{1+z}
\left[\frac{M}{M_{\sun}}\frac{18\pi^{2}}{\delta_{c}}\frac{\Omega_{m}(z)}{\Omega_{m}(0)}\right]^{1/3}
\end{eqnarray}
with $r_{0}=1.69\ {\rm Mpc}/(h p)$ where $p$ is the ratio between the virial to core radius 
$R_{v}/r_{c}$ (taken here to be 10).
The distribution of cluster bulk velocities is derived from the continuity 
equation 
\begin{eqnarray}
v_{{\rm rms}}^{2}=\int\frac{dk}{2\pi}P(k).
\end{eqnarray} 

The calculation of $\xi({\bf l})$ is significantly simplified 
by focusing on the {\it total} polarization which unlike $Q$ and $U$ - 
is a scalar field (i.e. a function of r only, as in Eq. 17 or the 
2D-projected $\beta$-profile), and - as in the case of temperature 
anisotropy - we can employ the expansion of scalar-valued 2D plane 
wave in cylindrical Bessel functions
\begin{eqnarray}
e^{i{\bf l}\cdot\theta}=\sum_{l}i^{l}J_{l}(l\theta)e^{i(\phi-\phi_{l})}
\end{eqnarray}
to obtain
\begin{eqnarray}
\tilde{\xi}({\bf l})=2\pi\int\theta d\theta J_{0}(l\theta)\xi(\theta)
\end{eqnarray}
where $\xi(\theta)$ is the total polarization. Also, use of a 
$\beta$-profile for the density implies 
$\xi(\theta)=\xi(\theta/\theta_{c})$, with $\theta_{c}$ having 
explicit $M$ and $z$ dependence 
$\theta_{c}=\theta_{c}(M;z)$ through the spherical collapse model.
Now the mass and the radius are related through the density which is 
fixed to be $18\pi^{2}\rho_{b}(z)$, where $\rho_{b}$ is the 
background density following the simple spherical collapse model. This 
density depends on the redshift and therefore the electron density 
depends on $r$, $M$ and $z$.

To obtain $\xi(\theta)$ we simply take the square root of $Q$ and 
$U$ added in quadrature at an angular distance
$\theta=r/d_{A}$ from the cluster, where $r$ and $d_{A}$ are the 
physical distance from the cluster center ($r=\sqrt{X^{2}+Y^{2}}$), 
and the angular diameter distance to the cluster, respectively.

\section{Results}

All power spectra presented in this work were calculated assuming the 
$\Lambda$CDM with 
the WMAP5 best-fit parameters (Komatsu et al. 2008): baryon, dark matter 
and dark energy densities (in units of the critical density) 
$\Omega_{b}h^{2}=0.0227$, $\Omega_{c}h^{2}=0.1099$, 
$\Omega_{\Lambda}=0.742$, respectively, and scalar spectral index, 
Hubble constant in units of 100 km/sec/Mpc $n_{s}=0.963$, $h=0.72$, 
respectively. The mass variance parameter is $\sigma_{8}=0.8$. 

\begin{figure*}
\epsfig{file=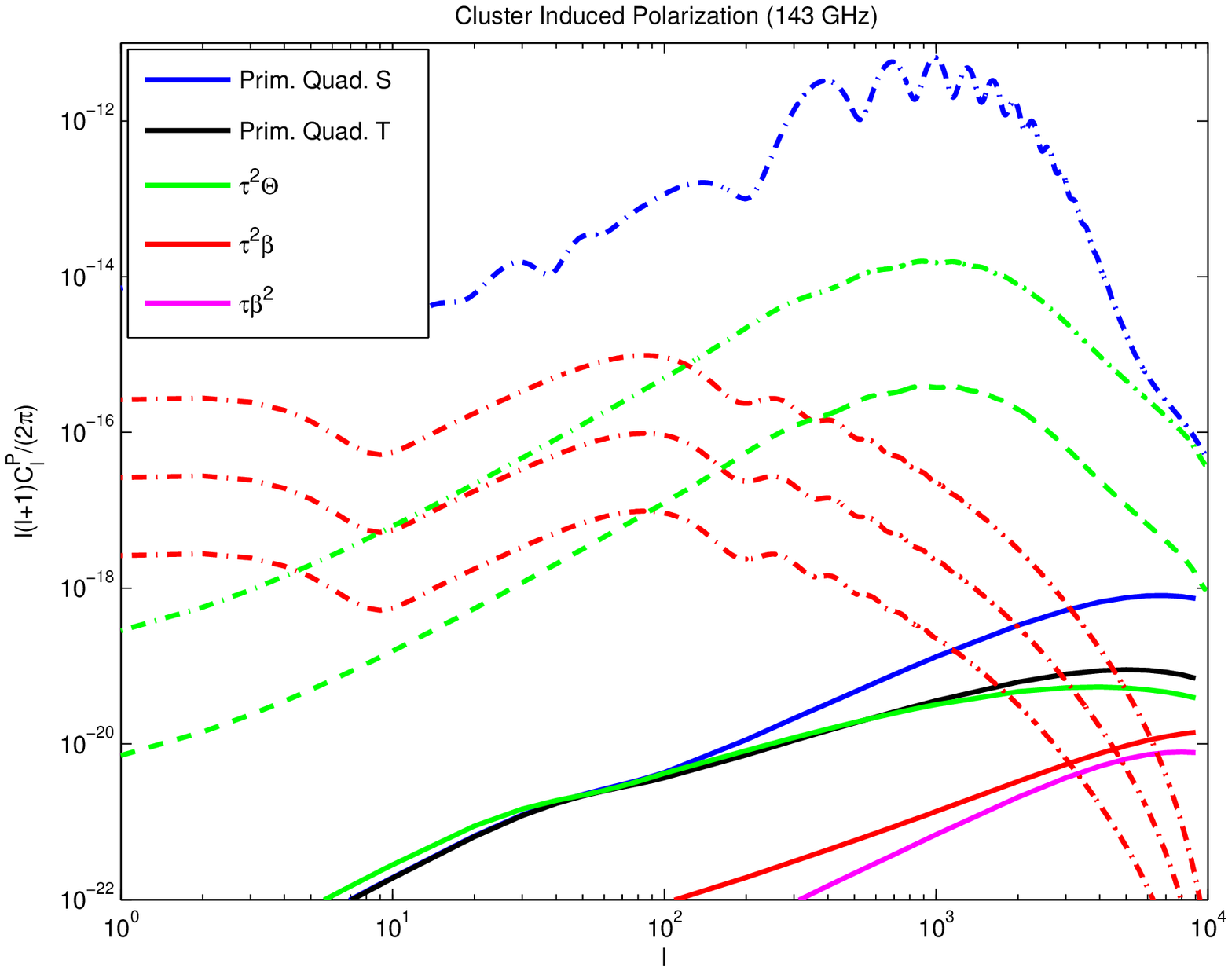, height=7cm, width=8cm,clip=}
\epsfig{file=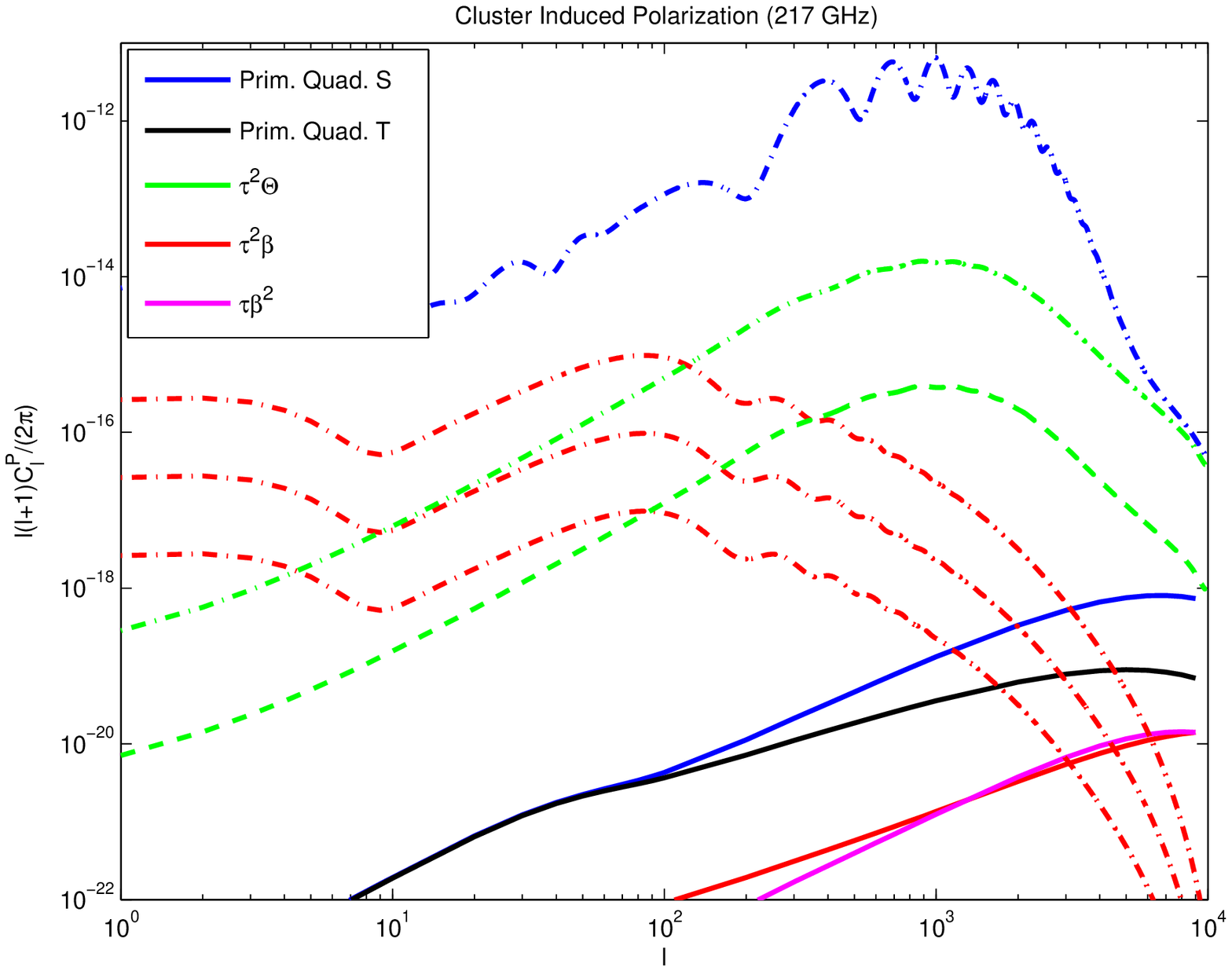, height=7cm, width=8cm,clip=}
\epsfig{file=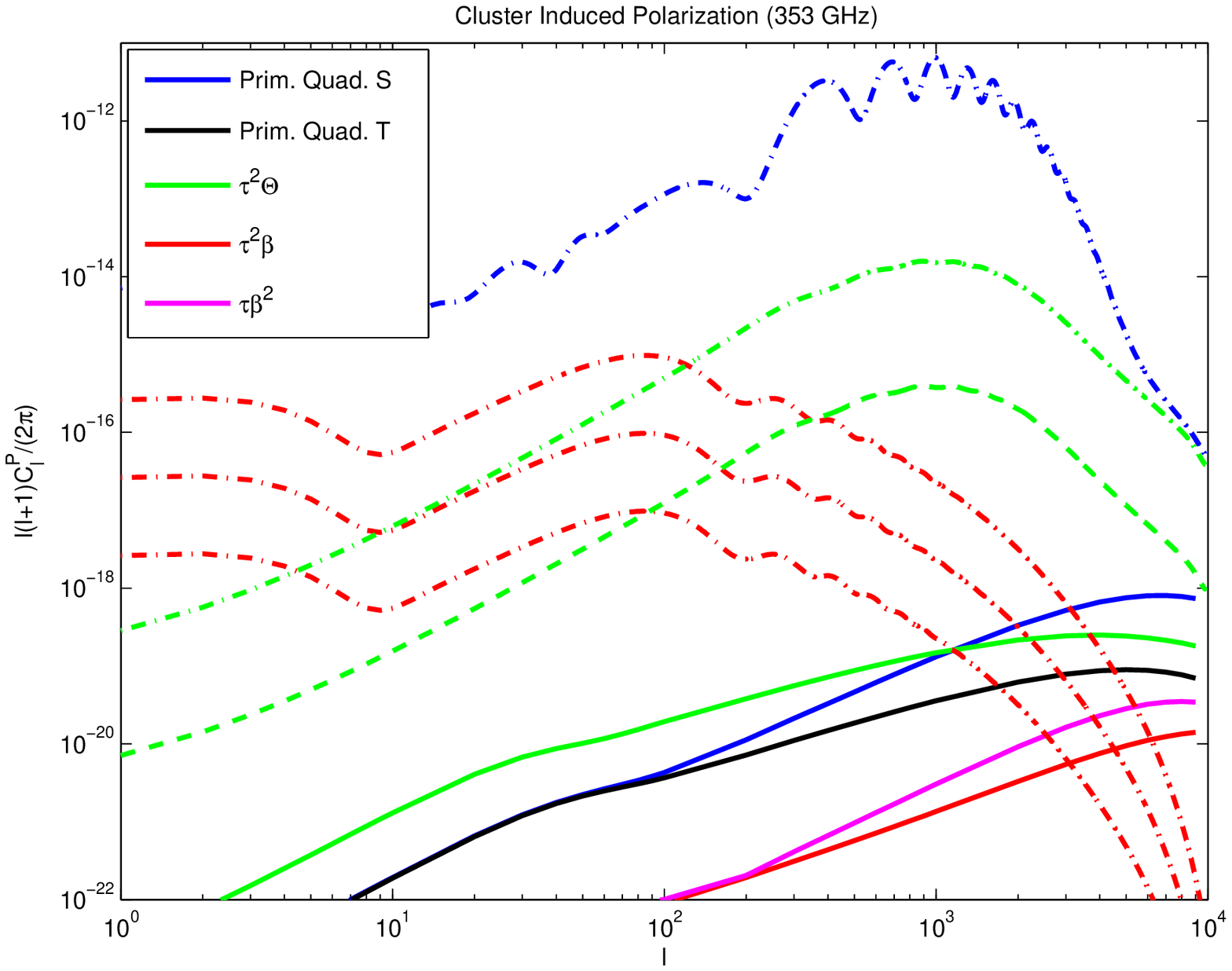, height=7cm, width=8cm,clip=}
\epsfig{file=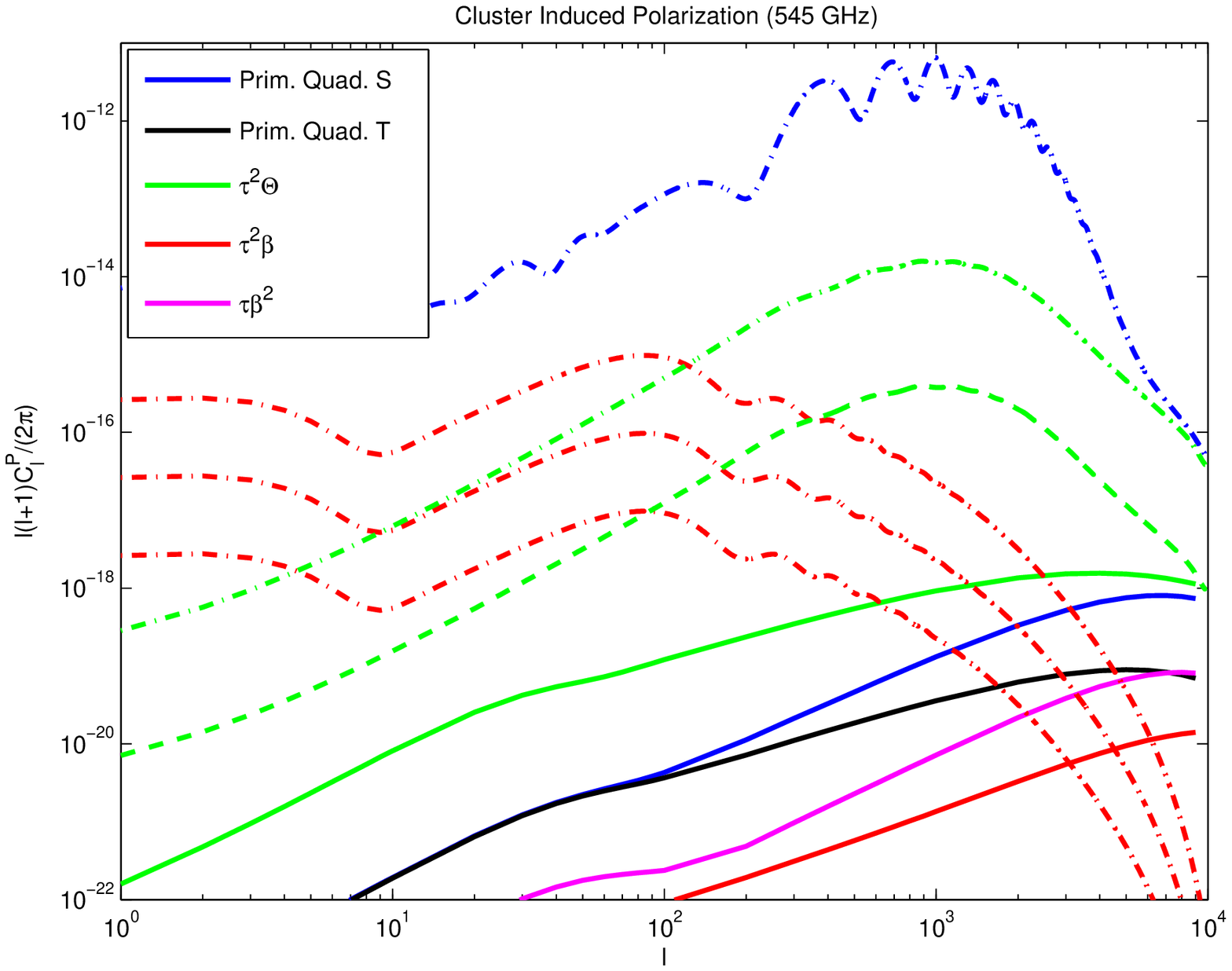, height=7cm, width=8cm,clip=}
\caption{SZ polarization at Planck HFI frequencies: Dot-dashed red 
lines are the B-modes from the primordial tensor perturbations with 
tensor-to-scalar-ratios $r=0.1$, $0.01$ and $0.001$ respectively. 
Dot-dashed blue curve is the E-mode polarization. Dot-dashed green 
curve is the B-mode from lensing conversion of the E-mode by the LSS. 
Assuming that this signal can be cleaned to 1 part in 40 
we plot the residual B-mode from lensing (dashed green curve).
Also shown are five SZ (total-) polarization signals: 
primordial scalar quadrupole (blue), upper limit (r=0.3) for the 
effect due to primordial tensor quadrupole (black), $\tau^{2}\Theta$ 
(green), $\tau^{2}\beta$ (red) and $\tau\beta^{2}$ (magenta).}
\end{figure*}

Power spectra of the cluster polarization components are shown in Fig. 2 
together with the corresponding ones of the primary anisotropy. The 
largest CMB polarized power is the primary E-mode polarization shown 
by the dot-dashed blue line in Fig. 2. Lensing of this E-mode anisotropy by 
the LSS produces B-mode pattern (dot-dashed green line in Figure 2) on 
a level which severely contaminates the primordial B-mode signal from 
inflation, shown by the dot-dash red lines in Figure 2 for three values 
of the tensor-to-scalar ratio $r$. As of yet the only information we 
have on $r$, which determines the amplitude of the B-mode from 
primordial gravitational waves, is an upper bound of $r<0.3$. Various 
techniques have been devised to remove the contamination due to the LSS 
from the primordial signal to allow the determination of the energy 
scale of inflation (e.g. Hu \& Okamoto 2002). 
However, below $r\approx 0.001$, the residual 
lensing still dominates the B-mode power even after the lensing 
contribution is optimally reduced by a projected factor of $\sim 40$, 
as demonstarated by the dashed green curve in Figure 2. Furthermore, 
for $r\leq 0.01$ the detection of the primordial B-mode is unlikely due 
to foreground contamination. Accordingly, the gravitational wave signals 
for $r=0.1$, $r=0.01$ and $r=0.001$ are shown for comparison.

Our results for the {\it total} polarization power spectra induced by 
scattering in clusters are shown by the solid lines in Fig. 2. 
Both the polarization due to scattering of the primordial quadrupole 
(scalar and tensor), the $\tau^{2}\beta_{T}$ effect, and the E/B 
conversion are independent of frequency, and thus are spectrally 
indistinguishable from the primordial signal. Separation between these 
contributions can possibly be based on their different statistical 
properties (e.g. the characteristic polarization pattern of the former, 
and non-gaussianity in the later case). Clearly, the largest B-mode 
signal comes from the E/B mixing by lensing. The $\tau\beta_{T}^{2}$ 
and the $\tau^{2}\Theta$ components do depend on frequency (as 
discussed in the previous section) and can be removed from the primary 
CMB signals by multifrequency observations. 

All our results are based on the $\Lambda$CDM with WMAP5 parameters. 
However, since the SZ effect is a sensitive function of matter 
clustering, $\sigma_{8}$, and since high-resolution CMB experiments, 
such as BIMA, CBI and ACBAR, which are sensitive to cluster angular 
scales infer higher $\sigma_{8}$ than the WMAP5 value of 
$\simeq 0.82$. With $\sigma_{8}=0.9$, as suggested by galaxy 
surveys, the cluster polarization power levels are higher by a factor 
of at least $\sim 2$ (see e.g. Sadeh, Rephaeli \& Silk 2007). 
Polarization from the tensorial quadrupole (black solid line in Fig. 2) 
is only an upper bound corresponding to $r=0.3$ (close to the current 
upper limit $0.22$).

\section{Discussion}

All cluster components calculated here are sub- to few nK and are smaller 
than the B-mode signal from lensing by $3-4$ orders of magnitude, too weak 
to have any noticeable impact on parameter estimation, even those which 
depend on the large scale structure, e.g. neutrino mass $M_{\nu}$, dark 
energy equation of state, $w$, etc. However, assuming that the lensing 
signal can be removed to 1 part in 40 by invoking optimal filters based 
on higher order statistics, cluster signals could potentially contaminate 
the {\it residual} lensing-induced B-modes signal on the few-percent level, 
again - too weak to have an impact on inflationary models.

Polarization signals from individual rich clusters may be detectable. 
While these signals may be negligible when averaged out on the full sky, 
current ground-based endeavors to detect the B-mode polarization 
typically target only few percent of the sky. Although these 
`radio-quiet' patches are optimally chosen to minimize the noise, it 
is not entirely excluded that CMB polarization by individual clusters 
will have an overall effect of more than the conservative few percent 
residual contamination shown in Figure 2. Also, it is important to 
reiterate here that the method of Hu \& Okamoto, mentioned above in the 
context of delensing the sky from the LSS-induced B-mode, indeed rests on 
the assumption that the unlensed signal is gaussian and that the 
expected small level of non-gaussianity is due to lensing by the LSS.
Therefore, it cannot be employed to remove the non-gaussian SZ 
polarization from clusters (which indeed was not calculated in this work). 
This will require, in principle, a customization of similar 
spatial-filtering methods to the case of galaxy clusters; a problem we 
have not addressed in this work, mainly because it is expected to be 
negligibly small for polarization. 

The morphology of IC gas typically shows a finite level of ellipticity. 
It is of interest to assess the implications of a small level of 
ellipticity on the small-scale power spectrum of the $\tau^{2}\Theta$ 
polarization component. As Figure 1 illustrates, the effect of double 
scattering by IC gas 
in a spherically symmetric cluster will result in vanishing polarization 
at the cluster center, due to the fact that no quadrupole is generated 
there from the first scattering. As a result, the $\tau^{2}\Theta$ component 
drops on very large $l$. However, cluster ellipticity changes this behavior 
because first scattering of photons travelling along the major and minor 
axes will result in a few percent quadrupolar change in their corresponding 
SZ-temperatures, and upon second scattering this small quadrupole will be 
polarized even at the cluster center. The leading order correction to the 
$\tau^{2}\Theta$ power spectrum will be $O(e^{2})$, rather than 
$O(e)$ (since both positive and negative ellipticity statistically 
avergae out to $\langle e\rangle=0$). In addition, the main qualitative effect will 
be to boost power at the cluster center as explained above, but 
this will probably be beyond the reach of even the highest resolution 
next generation experiments.

\section*{Acknowledments}

We acknowledge using CAMB to calculate the primordial power spectra.
BK gratefully acknowledges support from NSF PECASE Award AST-0548262.

\end{document}